\documentclass{ws-ijmpa}

\begin{document}

\markboth{W.Krzemie\'n et al. -- COSY-11 Collaboration}
{W.Krzemie\'n et al. -- COSY-11 Collaboration}

%
\catchline{}{}{}{}{}
%

\title{SEARCH FOR THE ${^3\mbox{He}}-\eta$ BOUND STATE AT COSY-11}

\author{\footnotesize W. Krzemie\'n$^{\star,\$}$$^,$\footnote{E-mail address: wojciech.krzemien@if.uj.edu.pl}~, 
J.~Smyrski$^{\star}$,
A.~Budzanowski$^{\#}$,
E.~Czerwi\'nski$^{\star,\$}$,
R.~Czy\.zykiewicz$^{\star}$,
D.~Gil$^{\star}$,
D.~Grzonka$^{\$}$,
L.~Jarczyk$^{\star}$,
B.~Kamys$^{\star}$,
A.~Khoukaz$^{\%}$,
P.~Klaja$^{\star}$,
P.~Moskal$^{\star,\$}$,
W.~Oelert$^{\$}$,
J.~Przerwa$^{\star}$,
J.~Ritman$^{\$}$,
T.~Sefzick$^{\$}$,
M.~Siemaszko$^{\&}$,
M.~Silarski$^{\star}$,
A.~T\"aschner$^{\%}$,
M.~Wolke$^{\$}$,
P.~W\"ustner$^{\$}$,
J.~Zdebik$^{\star}$,
M.~J.~Zieli{\'n}ski$^{\star}$,
W. Zipper$^{\&}$
}

\address{
$^{\star}$Institute of Physics, Jagiellonian University, 
Cracow, Poland\\ 
$^{\$}$IKP, ZEL, Forschungszentrum J\"ulich, J\"ulich, Germany\\ 
$^{\#}$Institute of Nuclear Physics, Cracow, Poland\\
$^{\%}$Institut f\"ur Kernphysik, Westf\"alische
Wilhelms-Universit\"at, M\"unster, Germany\\
$^{\&}$Institute of Physics, University of Silesia, 
Katowice, Poland\\ 
}

\maketitle


\begin{abstract}
We have measured excitation function for $dp \rightarrow ppp \pi^-$ reaction 
near the $\eta$ production threshold.
We observe an enhancement of the counting rate above the threshold which
can originate from the production of the $\eta$ meson in the reaction 
$dp \rightarrow {^3\mbox{He}}\,\eta$ and its subsequent absorption on
neutron in the $^3\mbox{He}$ nucleus leading to creation of the $p\pi^-$ pair.
\keywords{meson production; eta-mesic nucleus; final state interaction.}
\end{abstract}

\section{Introduction}  
Observation of a bound state of the $\eta$ meson and atomic nucleus would be very
interesting for studies of the $\eta-N$ interaction and for investigation 
of $N^*(1535)$ properties in nuclear matter. 
Existence of such states was postulated by Haider and Liu~\cite{haider},
however, up to now no firm experimental evidence for eta-mesic matter was found.
Encouraged by the recent data from MAMI showing some indications 
for photoproduction of ${\eta - ^3\mbox{He}}$ bound state\cite{pfei},
we performed a search for this state in the $d-p$ collisions.
The measurements reported here were carried out with the COSY-11 detector~\cite{cosy11}
using a slowly ramped internal deuteron beam 
of the COSY accelerator~\cite{cosy} scattered on a proton target~\cite{dombrowski}. 
The momentum of the deuteron beam was varied continuously within each acceleration cycle
from 3.095~GeV/c to 3.180~GeV/c, crossing the kinematical threshold 
for the $\eta$ production in the $dp \rightarrow {^3\mbox{He}}\,\eta$ reaction 
at 3.141~GeV/c~\cite{smyr1}.
A signature of existence of the $\eta-{^3\mbox{He}}$ bound state
would be an observation of a resonance-like structure with the center lying below
the $\eta$ production threshold in excitation curves for chosen decay channels. 
As decay modes we registered ${^3\mbox{He}}\,\pi^0$, $ppp \pi^-$ and $pd$ channel.\\
We measured also the total and differential cross sections for the $\eta$ production 
in the $dp \rightarrow {^3\mbox{He}}\,\eta$ reaction.
The cross sections from our measurements\cite{smyr1} 
and from similar measurements of the ANKE collaboration\cite{mers}
indicate a presence of a bound or virtual state in the $\eta-{^3\mbox{He}}$ system\cite{wilk}.
In turns, the excitation function for the  $dp \rightarrow {^3\mbox{He}}\,\pi^0$ process 
registered in the present experiment does not show any structure 
which could originate from the decay of $\eta-{^3\mbox{He}}$ bound state\cite{smyr2}. 
However, this does not exclude existence of such state since its signal 
can lie below the sensitivity of the present experiment due to small 
production cross section and/or due to low probability 
of decay in the registered ${^3\mbox{He}}\,\pi^0$ channel\cite{smyr2}. 
We expect that $ppp \pi^-$ can be a much more favorable decay channel
since it is produced in a one step process 
via absorption of the $\eta$ meson on the neutron
inside the $^3\mbox{He}$ nucleus leading to the creation of the $p \pi^-$ pair in the reaction 
$\eta n \rightarrow N^*(1535) \rightarrow p \pi^-$.
In the $\eta n$ center-of-mass frame, the pion and the proton 
are emitted back-to-back with momenta of about 431~MeV/c. 
In the reaction center-of-mass system
these momenta are smeared due to the Fermi motion of the neutron 
inside the $^3\mbox{He}$ nucleus. 
However, they are significantly larger than the momenta of the two remaining protons 
in the $^3\mbox{He}$ nucleus which play a role of "spectators"  moving
with the Fermi momenta which are in the order of 100~MeV/c.
In the next chapter we present recent results of analysis 
of the  $dp \rightarrow ppp \pi^-$ data collected in our experiment.

\section{Excitation function for the $dp \rightarrow ppp \pi^-$ reaction}

\begin{figure}[t]
\begin{center}
\begin{minipage}[t]{0.48\textwidth}
\centerline{\epsfysize 4.0 cm
\epsfbox{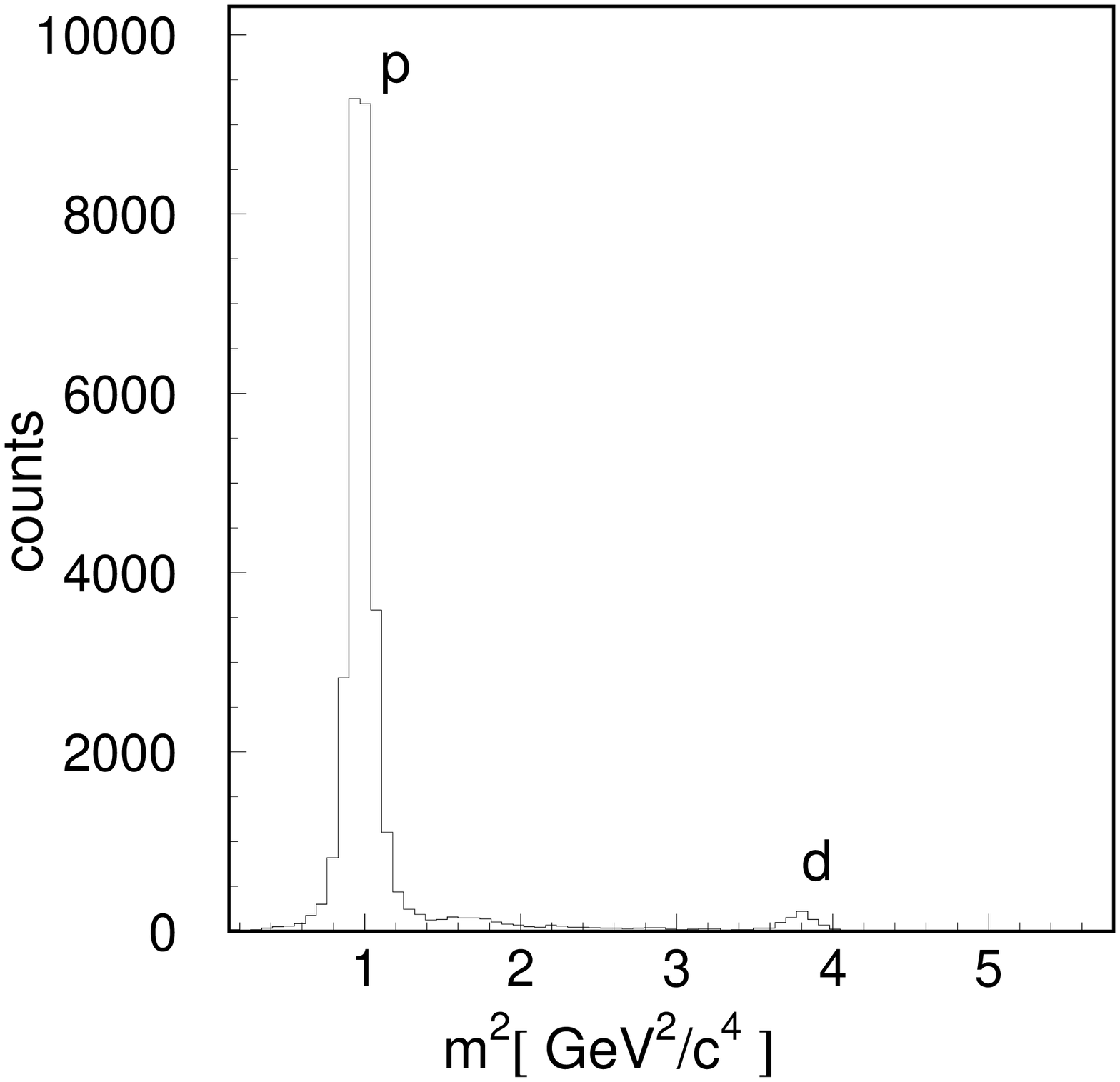}}
\caption{\label{fig1} Invariant mass squared determined 
on the basis of the time of flight between the S1 and S3 hodoscope.}
\end{minipage} \hfill
\begin{minipage}[t]{0.48\textwidth}
\centerline{\epsfysize 4.0 cm
\epsfbox{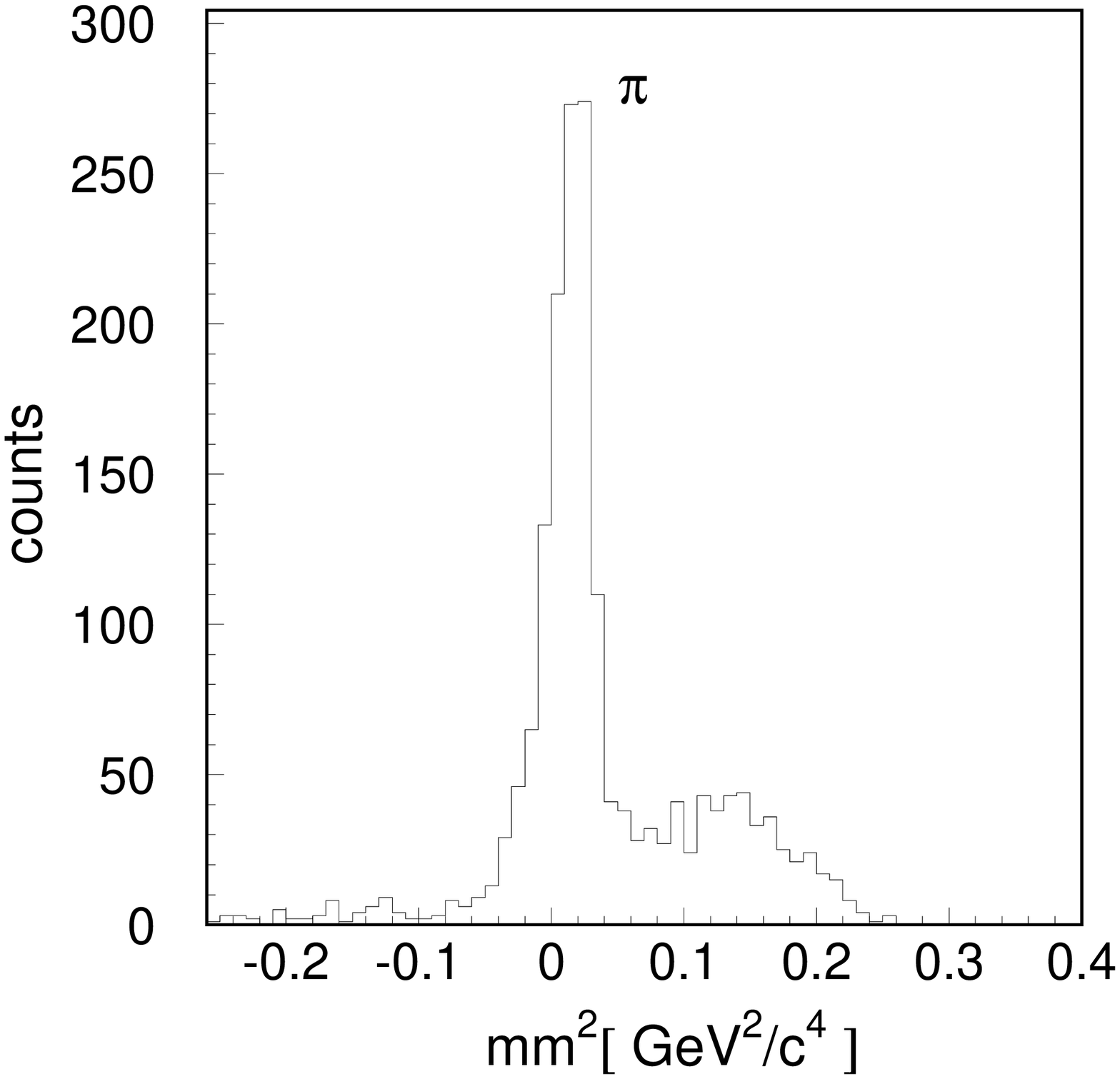}}
\caption{\label{fig2} Missing mass squared of the three protons system.}
\end{minipage}
\end{center}
\end{figure}

The protons originating from the $dp \rightarrow ppp \pi^-$ reaction
were momentum analyzed in the COSY-11 dipole magnet and their trajectories
were registered with a pair of drift chambers. 
A detailed description of the detection system can be found e.g. in articles~\cite{cosy11}.
The experimental trigger required at least three charged tracks in the scintillation 
hodoscope S1 standing in a distance of about 3~m from the target, and additionally at least two tracks 
in the hodoscope S3 placed 9~m behind S1. 
The momentum acceptance defined by the S1-S3 pair covered the momenta expected 
for the spectator protons.
The time of flight measured between S1 and S3 combined with the momentum 
measurement was used for identification of the outgoing protons (see Fig.~\ref{fig1}).
For identification of protons registered in the S1 but not reaching the S3 hodoscope, 
the time of flight between the target and S1 was used.
The pions were identified using the missing mass method (see Fig.~\ref{fig2}).

\begin{figure}[t]
\psfig{file=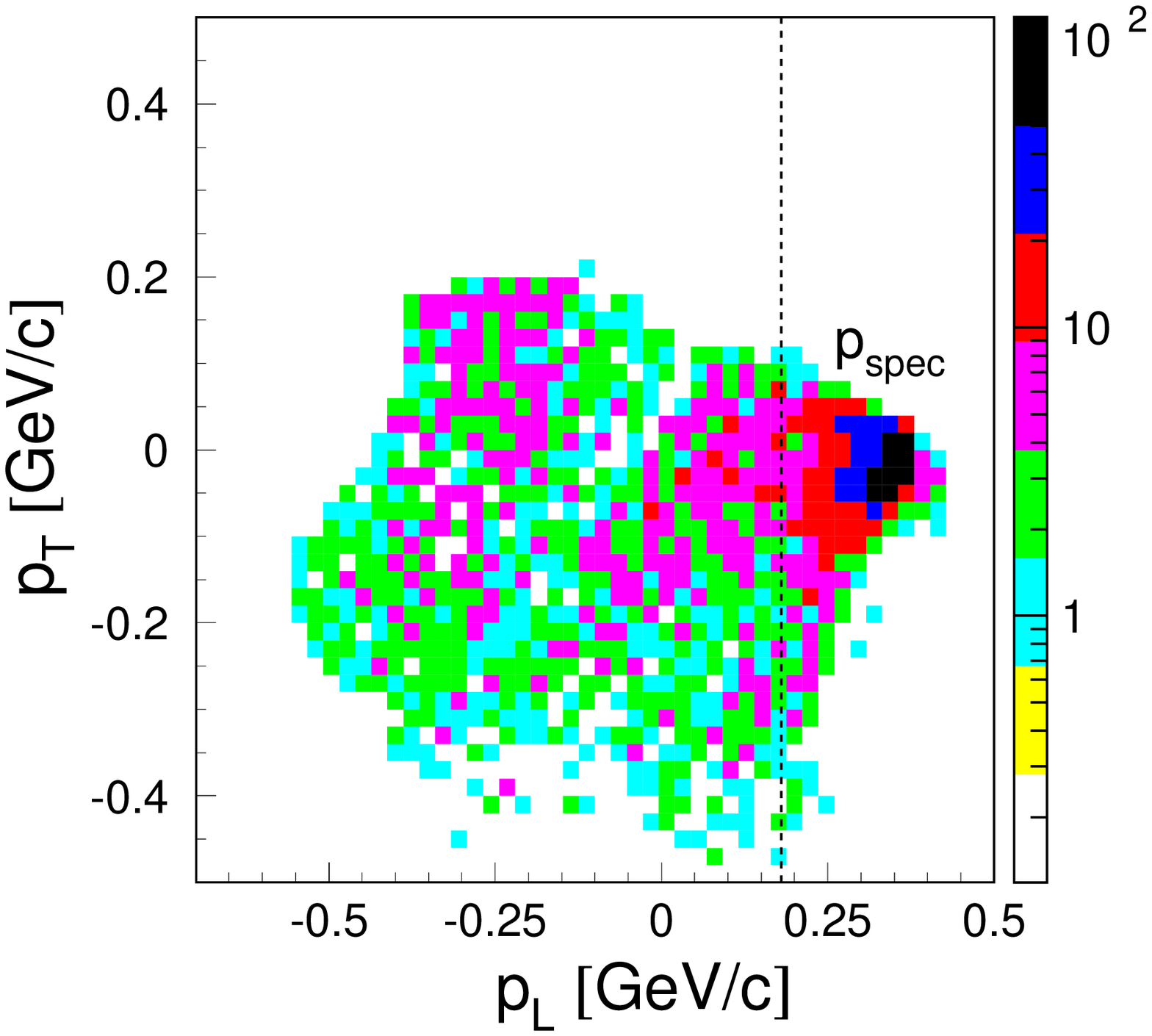,width=4.4cm}
\psfig{file=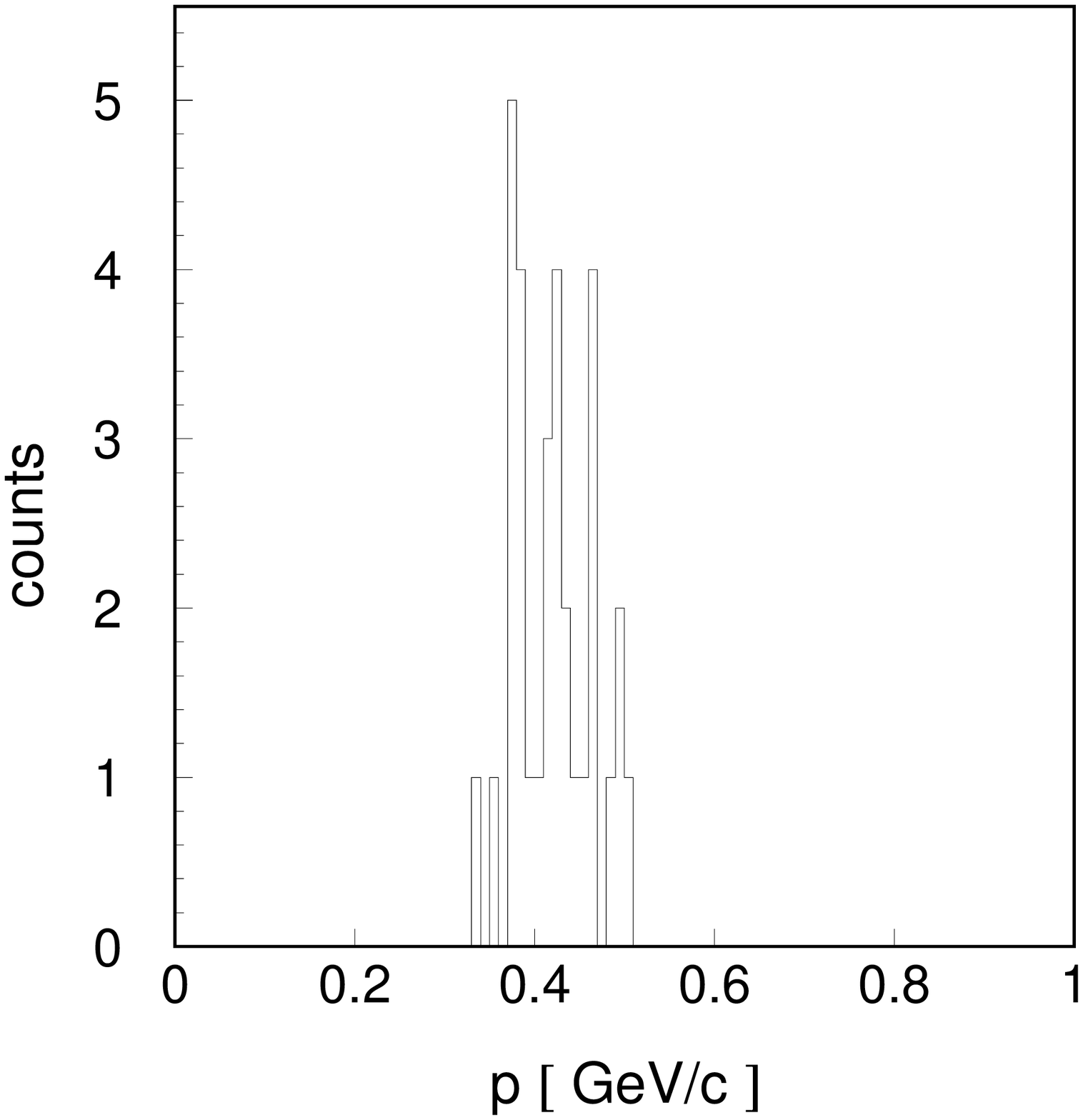,width=3.8cm}
\psfig{file=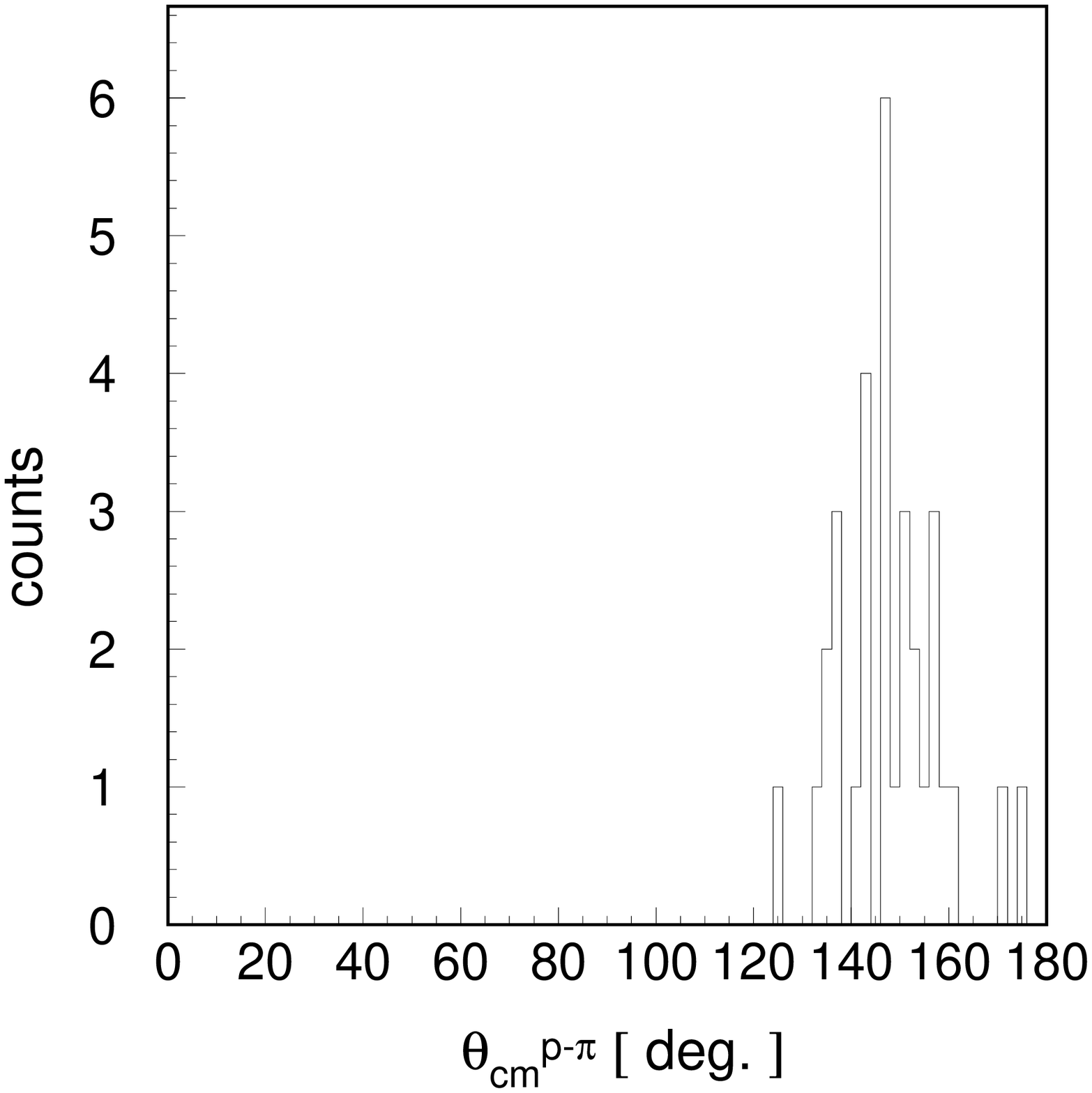,width=3.9cm}
\caption{\label{fig3} (left) Transversal vs. longitudinal momentum distribution 
of protons. The dashed vertical line represents the cut ($p_L < 0.18 GeV/c$)
applied on the longitudinal proton momenta. 
The counts at $p_L > 0.18 GeV/c$ correspond predominantly
to the spectator protons from the deuteron beam.
(middle) Distribution of pion momentum in the c.m. system.
(right) Angle between the pion momentum vector 
and momentum vector of the proton with the highest absolute value of c.m. momentum.}
\end{figure}

Left panel of Fig.~\ref{fig3} shows distribution of the transversal vs. longitudinal momentum 
components of registered protons from the $dp \rightarrow ppp \pi^-$ reaction.
This distribution is dominated by events of quasi-free
$\pi^-$ production in the process $np \rightarrow pp \pi^-$ where the neutron projectiles
originate from the deuteron beam.
The corresponding spectator protons from the deuteron beam are visible
as a group of counts on the right hand side of Fig.~\ref{fig3}(left).
In the further analysis we rejected the quasi-free $\pi^-$ production
by setting an upper limit for the longitudinal proton momenta equal to 0.18~GeV/c 
in the c.m. system, represented by the dashed line in Fig.~\ref{fig3}(left). 

The distribution of the $\pi^-$ momenta determined after application of this cut
is centered at the value of about 430~MeV/c (see middle panel of Fig.~\ref{fig3}) as expected
for pions originating from decay of the $N^*$(1535). 
Also the distribution of the c.m. angles between the pion momentum vector
and the momentum vector of the proton with the highest absolute value of the c.m. momentum
indicates for such decay since the angles are close to $180^{\circ}$(right panel of Fig.~\ref{fig3}).

\begin{figure}[t]
\begin{center}
\begin{minipage}[c]{0.40\linewidth}
\psfig{file=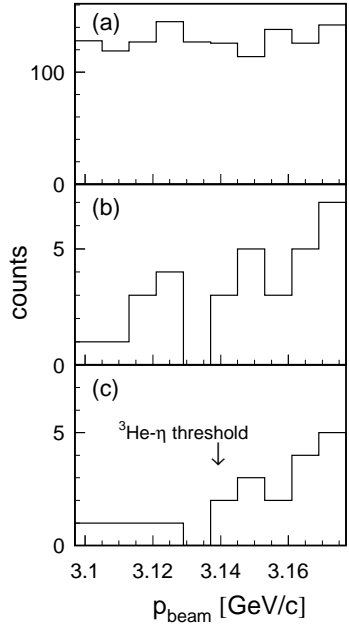, width=4.5cm}
\end{minipage}\hfill
\begin{minipage}[c]{0.56\linewidth}
\caption{\label{fig6} Number of $dp\rightarrow ppp\pi^-$ events as a function
of the beam momentum: without any cuts (a), after rejection of events
corresponding to the quasi-free $\pi^-$ production (b), and after additional
cut on the momenta of the spectator protons - $p^{cm}<200$~MeV/c (c).}
\end{minipage}
\end{center}
\end{figure}

The counting rate of all identified $dp\rightarrow ppp\pi^-$ events including the quasi-free
$\pi^-$ production remains constant in the scanned range of the beam momentum 
(see Fig.~\ref{fig6}a). However, after rejection of the quasi-free events,
the number of $dp\rightarrow ppp\pi^-$ counts in the beam momentum interval above
the $\eta$ threshold is higher than the number of counts in the beam momentum interval 
of equal width below the threshold (see Fig.~\ref{fig6}b). 
This difference is equal to $23-9=14$ and its statistical significance is of $2.5 \sigma$.
This difference only slightly decreases to the value of $16-4=12$ after application of 
an additional requirement that two out of the three outgoing protons have the c.m. momenta smaller than 200~MeV/c as expected with high probability for the spectator protons 
(see Fig.~\ref{fig6}c).
As a possible reaction mechanism explaining the observed excitation function we consider
production of the $\eta$ meson in the reaction $dp \rightarrow {^3\mbox{He}}\,\eta$
which subsequently convert to $\pi^-$ in the interaction with the neutron 
in the $^3\mbox{He}$ nucleus in the process $\eta n \rightarrow N^*(1535) \rightarrow p \pi^-$.

\section*{Acknowledgments}
We acknowledge the support of investigations by the
European Community-Research Infrastructure Activity
under the FP6 programme (Hadron Physics,
RII3-CT-2004-506078), by
the Polish Ministry of Science and Higher Education under grants
No. 3240/H03/2006/31  and 1202/DFG/2007/03 and 
by the German Research Foundation (DFG).

\end{document}